\documentclass{PoS}
\pdfoutput=1
\title{Status of the Large Size Telescopes of the Cherenkov Telescope Array}

\ShortTitle{Status of the CTA LST}

\author{\speaker{Juan Cortina} for the CTA LST
  project\footnote{for consortium list see PoS(ICRC2019)1177}\\
  CIEMAT, Avda. Complutense 40, Madrid E-28040 Spain\\
  E-mail: \email{juan.cortina@ciemat.es}}

\abstract{The Cherenkov Telescope Array (CTA) will consist of two arrays
  of Imaging Atmospheric Cherenkov
  Telescopes (IACTs) at the northern and southern hemispheres. CTA will feature IACTs
  with mirrors of three different sizes optimized to cover different energy ranges. The
  proposed sub-arrays of four Large Size Telescopes (LST) at CTA-North and CTA-South
  target the lowest energy range between $\sim$20 GeV and 100 GeV.
  Thanks to their low weight of $\sim$110 tons the LSTs can move by 180 deg in azimuth
  in 20 seconds for Gamma Ray Burst (GRB) follow-up. An LST has a tessellated parabolic mirror of
  23 m diameter equipped with a system of actuators to correct for gravity-induced
  deformations during data taking.
  Its low-weight 2 ton camera at the prime focus has a ~4.5 deg diameter, 1855 high QE PMTs and an
  embedded readout with 1 GSps sampling speed designed for data acquisition rates exceeding 10 kHz.
  A fully equipped LST has been installed at the CTA-North site in 2018 and is expected
  to be finished commissioning
  during 2019. The remaining three LSTs in the north will be installed by 2022. We will review the
  status of the LSTs, describe the installation of the first LST and report on the first results of
  the commissioning tests.}

\FullConference{36th International Cosmic Ray Conference -ICRC2019-\\
		July 24th - August 1st, 2019\\
		Madison, WI, U.S.A.}

\begin{document}

\section{Introduction}

CTA\cite{CTA} will consist of two arrays of IACTs at the northern and southern hemispheres. The IATCs 
will have mirrors of three different sizes optimised for different energy ranges.
The proposed
sub-arrays of four Large Size Telescopes (LST) at CTA-North and CTA-South feature the largest
reflectors and target the lowest energies down to a threshold energy of $\sim$20 GeV.
Major physics drivers in the LST energy range are transients, both galactic and extragalactic,
pulsars and studies of the Extragalactic Background Light. The design grants special attention 
to the study of GRBs: the telescope has a very low weight to allow repointing
by 180$^{\circ}$ in less than 20 seconds so as to detect the GRB prompt emission.

The team within CTA responsible for the development of the LST 
consists of more than 100 scientists from nine countries: Brazil, Croatia, France, Germany, 
India, Italy, Japan, Spain and Sweden\footnote{see a full institution list at
  {\it http://www.lst1.iac.es/collaboration.html}}. This team completed the design 
of the telescope in 2015. Development of the LST is different compared to the other CTA 
telescopes because the first LST (LST1)
constructed is also a prototype. However, this prototype will be fully functional, and once 
commissioning finishes and verification tests are successfully completed, demonstrating the LST 
fulfils all CTA requirements, LST1 will become the first production LST of CTA. 
This decision is motivated by the relatively small number of units to be produced and the elevated 
cost of each individual telescope. If during commissioning it is found that design modifications are 
needed, the first telescope will be retrofitted to ensure its performance  is equal to that of 
the other LSTs.

The second telescope will mark the beginning of the LST Construction phase. Several elements
manufactured in workshops of participating institutes for LST1 will be subcontracted to
industrial partners. Orders
will be timed based on the experience obtained during the construction of LST1, aiming for a peak
construction rate of two telescopes per year. 

\begin{figure}[!htb]
	\centering
		\includegraphics[width=0.8\textwidth]{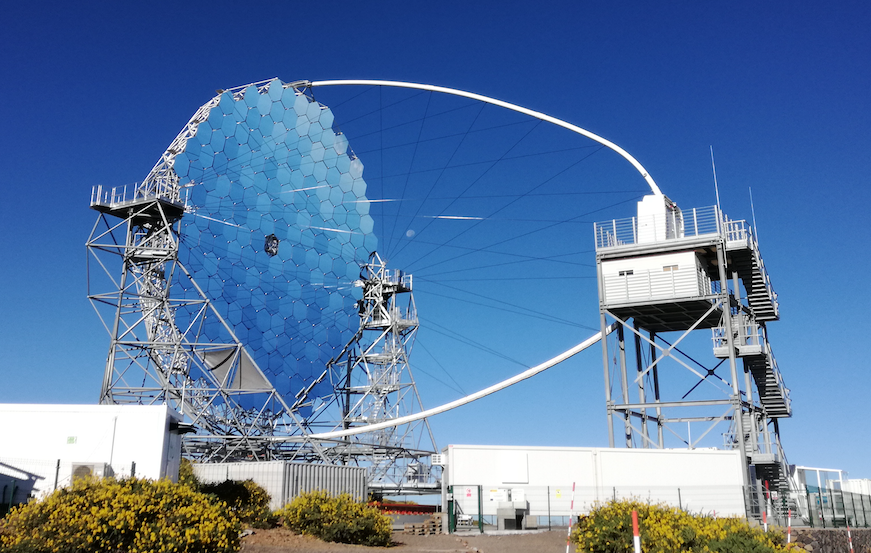}
	\caption{A picture of LST1 in June 2019.}
	\label{fig:lst_installation}
\end{figure}
The LST1 construction started at the CTA-North site with infrastructure and permission application in 2015. 
The foundation was built between July 2016 and January 2017.
The telescope itself was erected within 15 months between July 2017 and September 2018,
ending with the official inauguration of the telescope on October 10, 2018.
Figure \ref{fig:lst_installation} shows a picture of the telescope.

\begin{table}[!htb]
	\centering
	\begin{tabular}{| l | l | c | c |}
	\hline
		\multicolumn{4}{| l |}{\textbf{Optical Parameters}} \\
	\hline
				& Reflector type 			& 1-mirror, parabolic & 	\\
	\hline
	      	& Focal length 					& 28\,m 			& 	\\
	\hline
	      	& Dish diameter 				& 23\,m 			& 	\\
	\hline
	      	& f/D  				   & 1.2 			& 	\\
	\hline
	      	& Mirror area 		          & 396\,m$^{2}$	& 	w/o shadowing \\
	\hline
	      	& Mirror effective area     & 368\,m$^{2}$	& 	Including shadowing \\
	\hline
	      	& Preliminary on-axis PSF   & 0.05$^{\circ}$ & 	                \\
	\hline
	      	& Preliminary off-axis PSF   & 0.11$^{\circ}$ & at 1$^{\circ}$ off-axis 	      \\
	\hline
	      	& Preliminary tracking accuracy 	& 20\,arcsec 		& 	RMS, online precision \\
	\hline
	      	& Pointing accuracy 		& 14\,arcsec 		& 	RMS, post-calibration \\
	\hline
	         \multicolumn{4}{| l |}{\textbf{Camera Parameters}} \\
	\hline									
	      	& Camera dimensions (LxHxW)			& 	2.8\,m x 2.9\,m x 1.15\,m &  \\
	\hline
	      	& Weight                  			& 	$<$ 2000\,kg &  \\
	\hline
	      	& Number of pixels        			& 	1855  &  \\
	\hline
	      	& Pixel linear size       & 50 mm, incl light concentrator &  1.5\,inch PMT  \\
	\hline
	      	& Pixel field of view     			& 	0.1$^{\circ}$  &  \\
	\hline
	      	& Camera field of view     			& 	4.5$^{\circ}$  &  \\
	\hline
	      	& Trigger region field of view     			& 	4.5$^{\circ}$  &  \\
	\hline
	      	& Sampling speed           			& 	1 GS/s  &  \\
	\hline
	      	& Analogue buffer length   		& 	4\,$\mu$s  & for hardware stereo trigger  \\
	\hline
	      	& Readout rate            			& 	7.5\,kHz (required) 15\,kHz (goal) &  \\
	\hline
	      	& Dead time               			& 	5\% at 7.5 kHz &   \\
	\hline
	         \multicolumn{4}{| l |}{\textbf{Mechanical parameters}} \\
	\hline									
	      	& Total weight             			& 	103\,tons  & sum of moving parts  \\
	\hline
	      	& Repositioning speed      		& 	20\,s  & for 180$^{\circ}$ in azimuth  \\
	\hline
	      	& Zenith drive range      	     		& 	0$^{\circ}$ to 95$^{\circ}$ &   \\
	\hline
	      	& Azimuth drive range      	     		& 	408$^{\circ}$ &   \\
	\hline
	      	& Inertia elevation      	     		& 	$\sim$6000 tons$\cdot$m$^{2}$ &   \\
	\hline
	      	& Inertia azimuth       	     		& 	$\sim$12000 tons$\cdot$m$^{2}$ &   \\
	\hline
	      	& Park position       	     	  & 	zenith angle 95$^{\circ}$ &  locked at the camera tower \\
	\hline
	      	& Height at Camera Access        & 	13\,m above ground  &  In the parking position \\
	\hline		
	\end{tabular}
	\caption{Main LST parameters} 
	\label{tab:lst_params}
\end{table}
The reader is referred to the Technical Design Report (TDR) of the LST for 
a detailed description of the telescope and its component parts \cite{TDR}. 
The LST's main parameters are summarized in Table~\ref{tab:lst_params}.
The TDR also describes many of the tests that have been performed on the components
of LST1 before installation in La Palma.

\section{Infrastructure}

The LST1 foundation withstands all the loads transmitted by the bogies and the central axis in order to keep in place the telescope in any kind of situation. It is a circular construction made of concrete and iron bar armature. 
The foundation of the remaining telescopes will have a very similar design but it must be dimensioned for the actual sites of the observatory and for the specific conditions of the soil at the position of each telescope especially at the CTA-South site.
A circular rail of 23.9~m diameter and 500~mm width is fixed to the foundation by means of a pedestal, 
a curved I-beam. 
The steel feet are fixed with nuts to the concrete using special dowels that hold the required uplift forces.

Much of the basic infrastructure is currently installed in containers inside
the fenced area: an energy container equipped with two flywheels that deliver power to the telescope drive 
during fast repositioning, a drive container with drive control electronics, an IT container that is equipped with enough computing power for the whole CTA-North and space for most of the control computers and an office container where 
operators can control the telescope and monitor its status. Some of this infrastructure will be relocated to 
the CTA-North operations building, which should be finished in 2022. The remaining LSTs will only be equipped with a drive
and an energy container.

\section{Telescope structure and drive system}

The lower structure of the LST is made of steel tubes; the dish structure is a space-frame of carbon-fiber reinforced plastic (CFRP), steel and aluminium tubes. The lower structure rests on six wheel bogies running on the rail. 
\begin{figure}[!htb]
	\centering
		\includegraphics[width=0.8\textwidth]{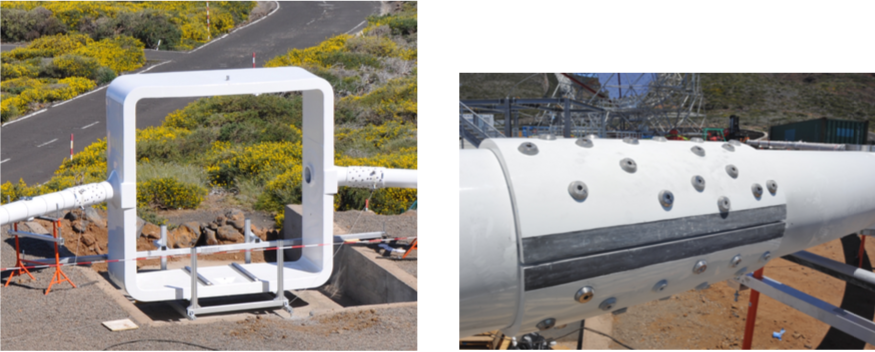}
	\caption{The CFRP camera frame during assembly with the CSS (left) and one of the junctions between 
		two sections of the CSS (right).}
	\label{fig:CSS_details}
\end{figure}
The baseline design of the camera support structure (CSS) is based on an almost parabolic arch
geometry, reinforced along its orthogonal projection by two symmetric sets of 
fixed headstays. Most of its elements make use of CFRP, which is well 
known to provide a very high performance to mass ratio.
On top of the CSS a square camera frame holds the camera at the 
focal distance. Figure \ref{fig:CSS_details} shows the camera frame and 
one of the CSS junctions.

To facilitate camera maintenance an access tower is permanently installed at the telescope parking position.
The camera is accessed from a flat platform on top of the access tower.

The fast and precise movement of the LST is achieved by using electric servomotors on
both the elevation and azimuth axis. Four synchronised motors are used for the azimuth axis and
two for the elevation one. 
There are two main strategies depending on the operational mode: fast motion or tracking. 
In fast motion, a regulation method limiting the torque is used in order to ensure a good load 
distribution between the motors. While in tracking mode the speed and not the torque of the motors is limited.
The azimuth and elevation motors are located on the top of the bogies and on the elevation drive arch, 
respectively. 

The park in/out of the telescope including access tower clamping and moving platform
is being automatized. More than 100 park in/out operations have been already performed.
The telescope moves already to target and starts tracking successfully. The drive
accuracy with respect to the rotation axis encoders is well below 30 arcsec.
Fast repositioning tests were performed in Spring 2019. The tests fully validated the 20s goal 
for 180 degrees in azimuth. Drive speed regulation worked as expected and the feedback from the mechanics
showed no issue. The emergency stops were successfully tested.

\section{Optics and active mirror control}

The optical system of the LST is an active optics system that includes a large parabolic reflector equipped 
with an active mirror control system (AMS) and a flat focal surface.
The optical reflector is composed of 198 hexagonal mirrors each with an area of 2~m$^2$. 
The mirrors are manufactured using the cold slump technique with a sandwich structure consisting of a soda-lime 
glass sheet, an aluminum honeycomb box and another glass sheet. 
The mirror box is made of stainless-steel.  Each mirror weighs about 47~kg.
The absolute mirror reflectivity between 300~nm and 550~nm 
is higher than 85\%. The optical PSF containment diameter ($\theta_{80}$) of a single facet is less
than 1/3 of a pixel size at the centre of the Camera.

The mirrors are attached to the dish of the LST structure using two actuators and a fixed point.
The actuators have accurate step motors (5~$\mu$m step size) which are 
controlled by the AMC program
to achieve the required optical performance at any moment of time. 
Each mirror facet has a small CMOS camera attached that observes a fixed-point (generated by a laser) 
on the Camera plane, and the position of the fixed-point is used as a reference to correct any 
misalignment of the mirrors. This active optics positioning will be done online with a frequency of 
0.1~Hz.

16 ''AMC boxes'' provide power to the actuators. The actuators are controlled by a standard industry PC 
installed in the AMC box. One box is responsible for 13 facets (26 actuators) at maximum. 
The 16 boxes are installed in the space frame of the telescope. The AMC software consists of two layers: a slave 
program controls each of the AMC boxes while a master OPCUA server controls the whole reflector. 
A master and a slave GUI have been developed. 

\begin{figure}[!htb]
	\centering
		\includegraphics[width=0.8\textwidth]{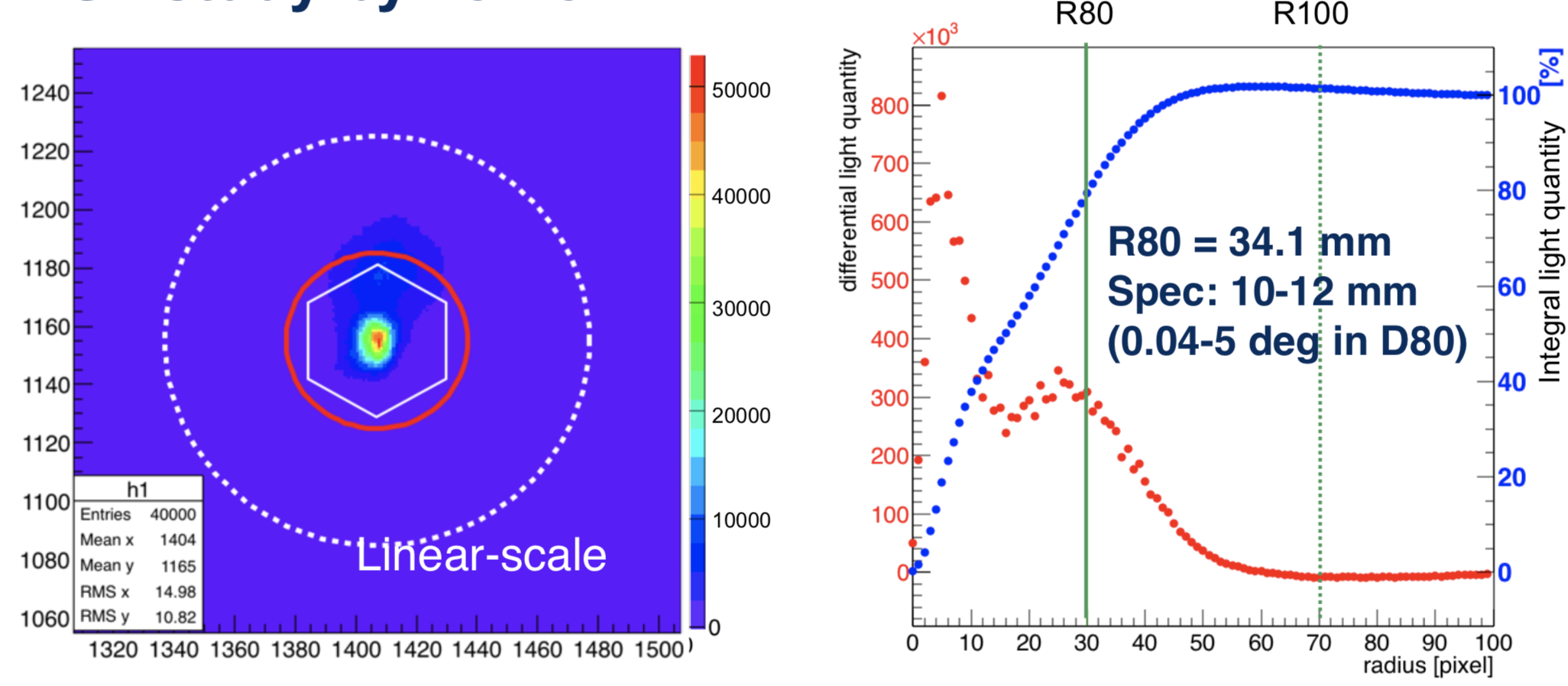}
	        \caption{Left panel: image of a star on the SIS. The hexagon corresponds to a pixel.
                  Right panel: differential/integral light flux
                  (in respectively a.u. or percentage) as a function of distance to the center
                  of the pixel. 80\% of the light is contained within 34.1 mm whereas the
                  requirement is 12 mm.}
	\label{fig:PSF_adjustment}
\end{figure}
First runs of the mirror adjustment process were completed: mirrors were adjusted group by group
on the shutter of the camera while the so-called star imaging screen (SIS) right in front of
the pixels was still 
unavailable. Mirrors were also adjusted one by one on the shutter, with an automatic image recognition.
Outliers were easily recognised and manually brought to the center of the PSF.
80\% of the light is already focused within a radius of 34.1~mm
(see figure \ref{fig:PSF_adjustment}). We believe that the final goal of 10~mm can be
accomplished once the final focus is performed.

The remaining hardware to complete the subsystem to determine the absolute pointing of the telescope
(reference laser for the AMC and starguider camera) will be installed in Summer 2019 \cite{pointing}. 

\section{Camera}

The camera of the LST has a weight of less than 2 tonnes and is equipped with 265 PMT modules 
that are easy to access and maintain. Each module has 7 channels. Hamamatsu photomultiplier tubes with a peak quantum efficiency of 42\% (R11920-100) are used as photosensors. 
Each photosensor is equipped with an optical light concentrator. 
The camera has been designed for maximum 
compactness and lowest weight, cost and power consumption while keeping optimal performance at low energies.

\begin{figure}[!htb]
	\centering
		\includegraphics[width=0.8\textwidth]{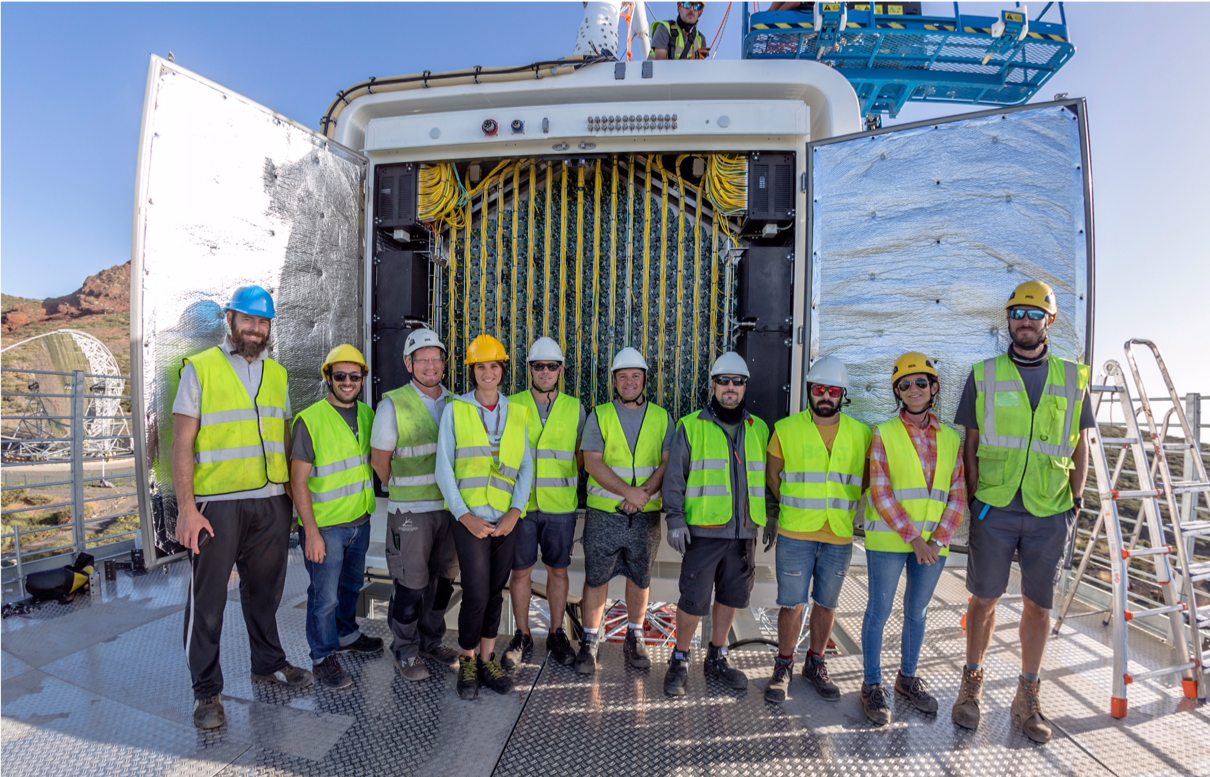}
	        \caption{Rear of the LST1 camera and a few members of the camera installation team in
                  September 2018. The team stands on top of a movable platform that opens
                  at night when the telescope starts observations.
	Most of the camera maintenance should take place through this rear door.}
	\label{fig:camera_back}
\end{figure}
Each pixel incorporates a photosensor as well as the corresponding readout and trigger electronics. 
This readout electronics is based on the DRS4 chip. 
In order to increase the analogue buffer length, 4 DRS4 channels are cascaded. The analogue
signals are split into Low and High gains. 
The camera trigger strategy is flexible and based on the
shower topology and the temporal evolution of the Cherenkov signal produced in
the camera. The analogue signals from the photosensors are conditioned and
processed by dedicated algorithms that look for extremely short and compact
light flashes. Furthermore, the cameras are interconnected in order to
form an on-line coincidence trigger amongst the LSTs. This enables the suppression
of accidental triggers by up to a factor of 100.

Manufacturing and assembly of the camera mechanics and electronics took several years
mainly at institutes in Japan and Spain (find details on some of the characterisation results
in \cite{camera}.). The mechanics were later assembled 
at the lab with the electronics and went through a round of exhaustive tests. The camera and PMT modules were
transported separately to La Palma, where they were re-integrated and tested again. The final installation into
the camera frame took only a few days in September 2018. Figure \ref{fig:camera_back} shows 
the camera right after installation.

\begin{figure}[!htb]
	\centering
		\includegraphics[width=0.8\textwidth]{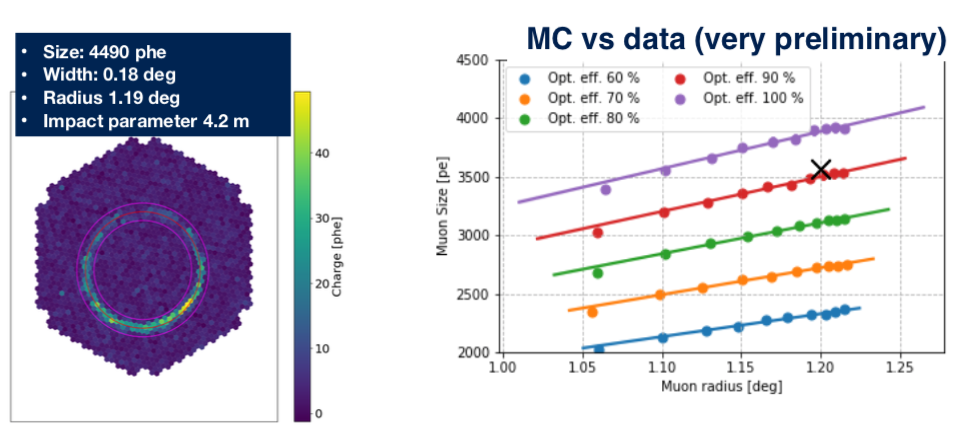}
	\caption{Left: an event containing a muon ring. Right: comparison with MC expectations. The size
	and radius would correspond to an optical efficiency of 90\%. This plot is very preliminary and only 
	meant to illustrate the calibration procedure. }
	\label{fig:muon_ring}
\end{figure}
During the last months the environmental variables inside the camera have remained within specifications 
both night and day. 
Many of the camera calibration procedures have been completed: trigger timing and PPS signal propagation,
and High Voltage flat fielding. The calibration of the trigger L0 settings (clipping, delay, attenuation) 
and L1 thresholds is ongoing.

Data taking runs on calibration light pulses and cosmic ray showers have been successfully taken. As an illustration
figure \ref{fig:muon_ring}, left, shows an event with a complete muon ring. The charge and radius of the
muon ring can actually be used to calibrate both the optical efficiency of the telescope and the optical PSF.
The right panel of the same figure shows how this specific muon ring would fit with the MC predictions.

The camera is calibrated with a 355~nm laser located at the center of the mirror dish. The laser was 
installed last year and is operational. Details can be found in \cite{calibration}.

A periodic trigger up to 18.8 kHz (with fixed gain selection) was reached. The rate is limited to
12 kHz without gain selection. For random triggers the highest possible rate so far is 11.5 kHz,
probably limited by TCP/IP congestion. This limit is under investigation.

\section{Outlook}

We expect to complete the commissioning of LST1 before the end of 2019. Observations with the telescope 
could then start although the performance of a single LST is significantly worse than the performance of the LST sub-array
or the whole CTA-North. The plans to complete the four LSTs at the CTA North site are firm.
The schedule of the production follows the funding flow in the countries with responsibilities in the
LST project. A good fraction of the components is already available, others are under production.
Installation of the telescopes is targeted before the end of 2022 and commissioning during
will take place during the next year. 

\section*{Acknowledgments}

This paper has gone through internal review by the CTA Consortium.

This work was conducted in the context of the CTA LST Project.

We gratefully acknowledge financial support from the agencies and organizations listed here: 
http://www.cta-observatory.org/consortium$\_$acknowledgments


\begin{thebibliography}{99}
\bibitem{pointing} 
D. Zaric for the CTA consortium,
\emph{Pointing System for the Large Size Telescopes Prototype of the Cherenkov Telescope Array} 
these proceedings.
\end{thebibliography}
\end{document}